\documentclass{article}
\usepackage{dcase2024,amsmath,graphicx,url,times,booktabs, tabularx, listings, xcolor}

\definecolor{codegreen}{rgb}{0,0.6,0}
\definecolor{codegray}{rgb}{0.5,0.5,0.5}
\definecolor{codepurple}{rgb}{0.58,0,0.82}
\definecolor{backcolour}{rgb}{0.95,0.95,0.92}

\makeatletter
\newcommand\codefontsize{\@setfontsize\codefontsize{6.5}{8.5}}
\makeatother

\lstdefinestyle{code}{
  backgroundcolor=\color{backcolour}, commentstyle=\color{codegreen},
  keywordstyle=\color{magenta},
  numberstyle=\tiny\color{codegray},
  stringstyle=\color{codepurple},
  basicstyle=\ttfamily\codefontsize,
  breakatwhitespace=false,         
  breaklines=true,                 
  captionpos=b,                    
  keepspaces=true,                 
  numbers=left,                    
  numbersep=5pt,                  
  showspaces=false,                
  showstringspaces=false,
  showtabs=false,                  
  tabsize=2
}

\title{SALT: Standardized Audio event Label Taxonomy}

\name{Paraskevas Stamatiadis, 
        Michel Olvera,
      Slim Essid
      }
\address{ LTCI, Télécom Paris, Institut Polytechnique de Paris, France  \\         
        \{paraskevas.stamatiadis, olvera, slim.essid\}@telecom-paris.fr\\
 }

\begin{document}

\ninept
\maketitle

\begin{sloppy}

\begin{abstract}
Machine listening systems often rely on fixed taxonomies to organize and label audio data, key for training and evaluating deep neural networks (DNNs) and other supervised algorithms. However, such taxonomies face significant constraints: they are composed of application-dependent predefined categories, which hinders the integration of new or varied sounds, and exhibits limited cross-dataset compatibility due to inconsistent labeling standards. To overcome these limitations, we introduce \textit{SALT: Standardized Audio event Label Taxonomy}. Building upon the hierarchical structure of AudioSet's ontology, our taxonomy extends and standardizes labels across 24 publicly available environmental sound datasets, allowing the mapping of class labels from diverse datasets to a unified system. Our proposal comes with a new Python package designed for navigating and utilizing this taxonomy, easing cross-dataset label searching and hierarchical exploration. Notably, our package allows effortless data aggregation from diverse sources, hence easy experimentation with combined datasets.

\end{abstract}

\begin{keywords}
Machine listening, DCASE, sound taxonomy, sound categorization, data aggregation
\end{keywords}

\section{Introduction}
\label{sec:intro}

Machine listening systems support a wide range of audio applications, including urban sound analysis \cite{vidana2021multilabel, angulo2023cosmopolite}, industrial acoustic monitoring \cite{suefusa2020anomalous, wichern2021anomalous, liu2022anomalous}, music analysis \cite{durrieu2008singer, ni2012end, oramas2018multimodal}, and speech recognition \cite{toshniwal2018multilingual, pratap2019wav2letter++, palaz2015convolutional}. The key success of these systems, typically relying on supervised machine learning approaches, especially using deep neural networks (DNNs), lies in the systematic annotation of training data, using predefined class labels from hierarchical sound ontologies and taxonomies \cite{salamon2014dataset, gemmeke2017audio, fonseca2021fsd50k, cartwright2019sonyc, zinemanas2019mavd, ooi2021strongly, turpault2019sound}. 

Sound ontologies and taxonomies serve as foundational frameworks to categorize everyday sound scenes and events \cite{guastavino2018everyday}. Developed across several research fields—for instance auditory cognition \cite{gaver1993world, guastavino2007categorization}, soundscape research \cite{schafer1994our}, sound design \cite{moffat2017unsupervised}—they are particularly instrumental in machine listening. Notable examples stand out for their wide adoption in the DCASE\footnote{Detection and Classification of Acoustic Scenes and Events} community: UrbanSound8K \cite{salamon2014dataset}, SONYC-UST \cite{cartwright2019sonyc, Cartwright2020SONYCUSTV2AU}, MAVD-traffic \cite{zinemanas2019mavd} for urban sound analysis and ESC-50 \cite{piczak2015esc} and AudioSet \cite{cartwright2019sonyc} for broader sound event recognition.

As evident from the previous examples, categorization of sounds in these systems are context-specific and tailored to desired applications. Their static nature often entails significant, overhauls for updates or extensions, especially when combining audio events from different environments, even for the same application. An exemplary case of this, is the adaptation of the SONYC-UST's taxonomy. Originally developed to classify urban sounds in New York City, this taxonomy required an expansion to accommodate the unique sounds of Singapore city's soundscapes, while maintaining compatibility with the base categorization. This adaptation allowed for bench-marking of urban sound tagging systems across different cities \cite{ooi2021strongly, 10095833}.

While recent initiatives such as \textit{mirdata} \cite{bittner2019mirdata} and \textit{Soundata} \cite{fuentes2021soundata} have simplified the use of major datasets for Music Information Research (MIR) and DCASE, by standardizing data loading, these efforts primarily focus on addressing issues related to data management and accessibility through open-source software packages. As such, these packages promote reproducibility and flexible data-processing pipelines. However, the development of adaptable and extensive sound categorization frameworks capable of integrating new audio event labels from diverse datasets while maintaining compatibility with existing taxonomies remains largely unaddressed.

In this work, we tackle such challenges by introducing \textit{SALT: a Standardized Audio event Label Taxonomy}. Leveraging the hierarchical structure of AudioSet, \textit{SALT} extends and standardizes labels across 24 publicly available environmental sound datasets. Such a large collection of datasets covers diverse audio analysis tasks including audio tagging, sound event detection and acoustic scene classification. By standardizing labels, SALT enables mapping them across diverse datasets, ensuring compatibility and easing dataset aggregation. Alongside our proposed taxonomy of standard dataset labels, we present \textit{py-salt}, an open-source python package designed to navigate through its content. This tool allows users to easily navigate through the hierarchical label taxonomy at any level of granularity. It turns out to be quite valuable when performing experiments considering various existing datasets whose particular labelling schemes can be seamlessly represented in our unified taxonomy.

We posit that our contribution is timely in a research context where large-scale training of audio models is fueled by the availability of (labelled) training data and computational resources. Our taxonomy with standardized audio event labels simplifies data aggregation, complementing tools like \textit{Soundata}  to develop audio classification models at scale.

The remainder of this work is organized as follows: Section \ref{sec:taxonomy}, introduces the motivation and design principles behind SALT. Section \ref{section:functionalities} presents the functionalities and applications of py-salt, our proposed Python package, and Section \ref{section:conclusion} concludes the article.

\section{SALT}
\label{sec:taxonomy}

The motivation behind creating SALT is the development of a new solution leveraging existing taxonomies to facilitate experimentation across different environmental sound datasets. The key feature of this solution is label aggregation, which allows unified categorization of sound events. This approach necessitates a standardized set of labels applicable to multiple environmental sound datasets. Consequently, with SALT we aim to expand AudioSet, the largest general-purpose sound event taxonomy, and use it as a common frame of reference to represent the annotations of all major publicly available DCASE datasets.

\subsection{Design Principles}

We aim to establish a general-purpose sound taxonomy with label aggregation capabilities at the core of its design. To achieve this, we adapt existing  sound event taxonomies from diverse domains, including but not limited to urban sound analysis, acoustic scene classification, domestic sound event detection, among others, using AudioSet's taxonomy as our basis. A key principle is to integrate labels from diverse sound collections, prioritizing datasets that are independent from each other rather than subsets of others, leading to a natural expansion of AudioSet's taxonomy. This integration into a unified taxonomy entails a standardization process to ensure label consistency across datasets.
\\\\
\textbf{Label standardization}. The standardization process involves a mapping of original (\textit{i.e}., default) category names from different datasets that describe the same acoustic event to a standardized label. For example, labels such as  ``car horn'' in \textit{UrbanSound8K}, ``car\_horn'' in \textit{ESC-50} and ``Vehicle horn, car horn, honking" in \textit{AudioSet}, all refer to \text{the sound produced by a car horn}. To aggregate labels effectively, we map them to the standard label \textit{car\_horn} in SALT. Our notation for denoting standard labels uses lowercase characters and underscores instead of white spaces.
\\\\
\textbf{Mapping for accurate aggregation}. In cases where a dataset label indicates more than one acoustic event, or sources producing sound, the mapping depends on the nature of the sounds. If the events or sources have similar acoustic properties, the word ``or'' is introduced in the standard label to preserve both sounds in the label. For example, the label ``Railroad car, train wagon'' in AudioSet is mapped to the standard label \textit{railroad\_car\_or\_train\_wagon} as both sources produce the same type of sound. On the contrary, when a dataset label indicates multiple sound events, each of them entailing unique acoustic signatures, the mapping selects the most specific (\textit{i.e.,}finest-grained) standard label that avoids incorrect associations in the aggregation process. For example, the label ``dog-barking-whining'' in \textit{SONYC-UST}  is mapped to the broader standard label \textit{dog} to ensure accurate aggregation. This principle prevents mistakenly including unrelated events into more specific standard labels such as \textit{dog\_barking} or \textit{dog\_whining}. In Figure \ref{fig:overview_salt} we present a clear depiction of our label standardization procedure. 
\\\\
\textbf{Hierarchy expansion}. Additionally, our objective is to preserve the base hierarchy of AudioSet while integrating new standard labels when strictly necessary. This design principle serves two main purposes. First, it facilitates label aggregation across multiple hierarchical levels by mapping dataset labels not only to a standard label, but also to its hierarchical ancestors (also standardized labels). For example, the label ``Bird'' in AudioSet is mapped to the standard label \textit{bird} in our taxonomy, as well as to its standard ancestors \textit{wild\_animal} and \textit{animal}. Second, it refines the AudioSet taxonomy by incorporating new or rare sound event labels coming from a wide variety of environmental sound datasets serving different audio analysis tasks. When a dataset contains class labels which do not fit neatly into the AudioSet taxonomy or cannot be covered by any existing node in the structure, new standardized labels are introduced to accomodate such labels. For instance, labels such as ``truck/compressor'' from the \textit{MAVD-traffic} and ``Friction brake'' from the \textit{SINGA:PURA} datasets, represent cases where AudioSet's existing labels are insufficient to fully capture the diverse set of sounds encountered in publicly available datasets.

\begin{figure}[t]
  \centering
  \centerline{\includegraphics[width=\columnwidth]{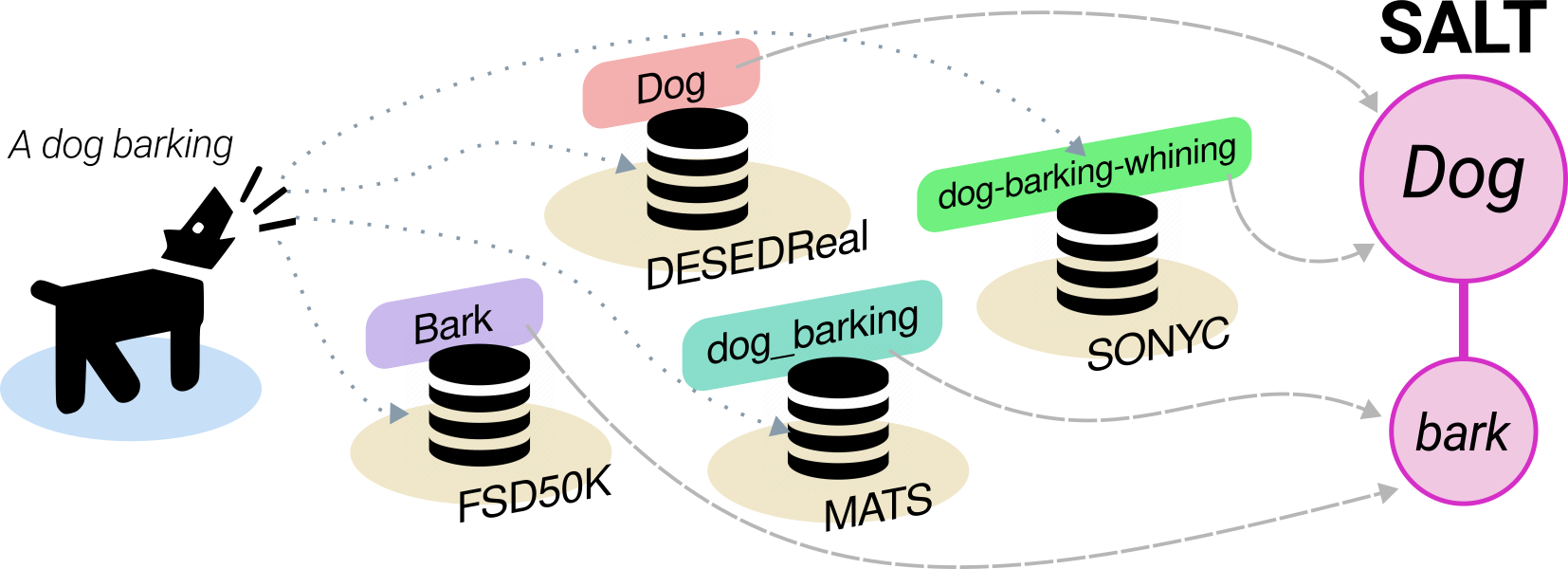}}
  \caption{Illustration of SALT's standardization process. Dataset labels are systematically mapped to a standard label that ensures cross-dataset compatibility.}
  \label{fig:overview_salt}
\end{figure}

\subsection{Taxonomy Structure}

Our proposed extension to Audioset's taxonomy, is structured into multiple hierarchical levels, each representing a different granularity of sound categories. Starting from AudioSet's seven broad sound categories — ``Human sounds'', ``Animal'', ``Music'', ``Source-ambiguous sounds'', ``Sounds of things'', ``Natural sounds'', and ``Channel, environment and background'' — and 616 sound labels (out of the 632 provided in the taxonomy), we expand to 734 sound labels. These labels are categorized under the original seven AudioSet categories, with the addition of two new categories: \textit{Water} and \textit{Other}. Figure \ref{fig:pie_graph} illustrates the contribution of the original labels from all considered datasets to compose the standard labels in SALT.

With careful examination of the video clips available in AudioSet and their associated labels, we refined the hierarchical structure of AudioSet to clarify the placement of labels within the taxonomy. For example, when examining the label ``Water'', we found that 37\% of clips include tags related to water sounds occurring in domestic environments \textit{e.g., ``Water faucet''}, while, only 2\% of them are related to outdoor and/or natural landscapes. This distribution indicates that the ``Water'' tag does not exclusively belong under ``Natural sounds'', but also frequently appears in domestic settings. Additionally, we conducted refinements by examining children within categories such as ``Vehicle'' and ``Engine''. For example, clips tagged with the label ``Accelerating, revving, vroom'', categorized under ``Engine'', primarily pertain to vehicle sounds, accounting for approximately 93\% of its instances. Therefore, ``Accelerating, revving, vroom'' is additionally categorized under ``Vehicle''. For a complete list of all such refinements, we refer the reader to our companion repository\footnote{\url{https://github.com/tpt-adasp/salt}}.

\begin{figure}[t]
  \centering
  \includegraphics[width=1\columnwidth]{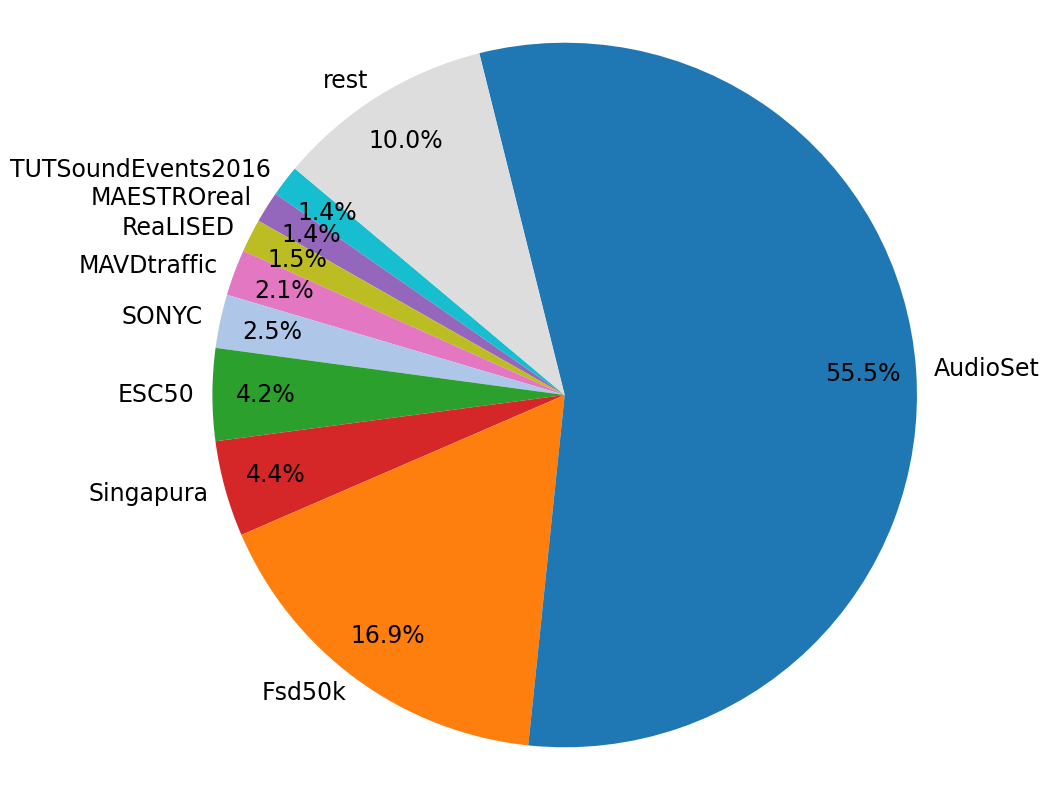}
  \caption{Contribution of dataset's original (default) labels to SALT after the standardization process.}
  \label{fig:pie_graph}
\end{figure}

\section{py-salt}
\label{section:functionalities}

In this section, we give an overview of the functionalities and applications of SALT, designed to unify event labels through standard labels, allowing for label search, data exploration, and hierarchical parsing. To exploit the benefits of our proposed taxonomy with standard labels, we developed \textit{py-salt}, a Python package that provides tools for navigating and utilizing the taxonomy.

\subsection{Functionalities}

\textbf{Label searching}.  This functionality allows two searching modes. First, standard labels can be employed to look for corresponding original (default) dataset labels across all those integrated in SALT, \textit{e.g., motorcycle}  \(\rightarrow\) ``motorbike'' coming from \textit{Urbansas}, ``Motorcycle'' from \textit{AudioSet/FSD50K}, ``motorcycle/wheel\_rolling'', ``motorcycle/engine\_idling'', ``motorcycle/engine\_accelerating'' from \textit{MAVD-traffic}, etc.
Secondly, original dataset labels can be employed to identify their counterparts across all datasets within the taxonomy. This dual approach offers comprehensive coverage and consistency for cross-dataset retrieval, \textit{e.g., ReaLISED's} ``water tap'' \(\rightarrow\) ``Water\_tap\_and\_faucet'', ``Water tap, faucet'', ``water tap running'' coming from \textit{FSD50K}, \textit{AudioSet} and \textit{TUT Sound Events 2016}, respectively.
\\

\noindent\textbf{Hierarchical exploration and expansion}. This functionality allows browsing the taxonomy at any level of the hierarchy and easily locate superodinate (parent), subordinate (child) and coordinate (sibling) categories. Additionally, SALT supports \textit{mapping expansion}, a functionality useful to incorporate new datasets and label categories into the taxonomy. The mapping process can be performed using the existing standard labels or by defining new ones suiting the user's requirements.
\\ \\
\textbf{Visualization and searching tools}. Graph plotting utilities are included in \texttt{py-salt}, which allows users to explore SALT visually. The python library contains methods to plot graphs showing the hierarchical structure of a given standard label in SALT, and also to depict all original (default) class names in the aggregated datasets that mapped to a SALT label. For example, the function \verb|plot_hierarchical_tree_graph('bird')| serves to generate a graphical representation of the hierarchical structure for the standard label \textit{bird} as illustrated in Figure \ref{fig:std_label_mapping}. This functionality provides a clearer depiction of the relationships between parent, child and sibling categories. Similarly, the function \verb|plot_std_label_mapping('car_horn')| serves to show the mapping of dataset labels to the standard label \textit{car\_horn} as illustrated in Figure \ref{fig:dataset_label_mapping}. This is useful for identifying the potential datasets needed for a specific application of interest. 

An additional example, is illustrated in Listing \ref{lst:code_example}, which involves retrieving the dataset labels mapped to the standard label \textit{reverse\_beeper}. The function returns a Python dictionary where dataset names serve as keys and their corresponding labels as values. This example highlights the package's feature to provide detailed and well-organized information about class labels, which is essential for analysis and integration of data.  

Furthermore, the package comes with extensive documentation including a tutorial notebook and practical examples to demonstrate all functionalities discussed in this section. The interested reader is referred to the corresponding repository for more information about \texttt{py-salt}.
\\

\begin{figure}[t]
  \centering
  \centerline{\includegraphics[width=\columnwidth]{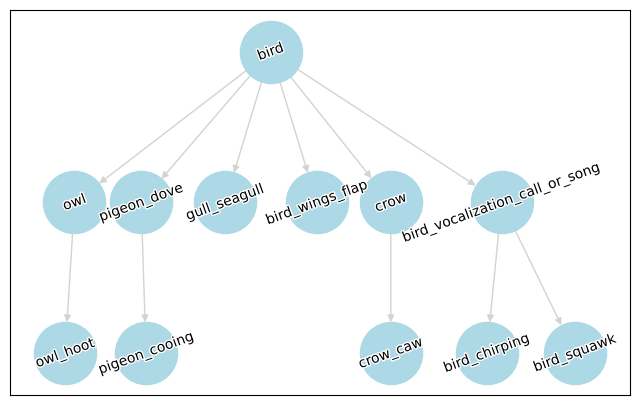}}
  \caption{Example of standard label mapping for the standardized label \textit{bird}.}
  \label{fig:std_label_mapping}
\end{figure}

\subsection{Applications}

SALT can serve diverse applications and use cases through its functionalities. The provided python library, facilitates exploration of mapped datasets, both individually and in combination. 
An interesting use case comprises the compilation of data from multiple datasets to compose new datasets or collections. This is achieved by the use of a series of methods provided in \texttt{py-salt}, that allows gathering the desired labels from specific datasets or domains of interest. For example, to develop an audio classifier specialized in the detection of emergency signals, all relevant labels from different datasets \textit{e.g., AudioSet, SINGA:PURA, ESC-50, etc.} can be easily accessed through the standard label \textit{alarm\_signal}. Similarly, to develop an urban sound monitoring system, various dataset labels can be aggregated through a set of standard labels such as \textit{vehicle}, \textit{engine}, and \textit{outdoor\_urban\_or\_manmade} from datasets \textit{MAVD-traffic, AudioSet, FSD50K} and \textit{SONYC-UST}, respectively. To give another example, for a system targeting the detection of domestic sound events, labels such as \textit{kitchen} (\textit{i.e. kitchen sounds}), \textit{bell} and \textit{television} can be aggregated to create a specialized classifier for recognizing common household sounds. Figure \ref{fig:statistics} illustrates the significant benefits of label aggregation in augmenting the amount of data available for minority classes.

Another use case involves defining a common set of standard labels for cross-dataset evaluation purposes, \textit{e.g}., training on \textit{SONYC} and testing on the same set of labels on \textit{UrbanSound8K}. This approach is particularly useful for bench-marking audio analysis systems and assess their generalization capabilities. Overall, SALT can diminish inconsistencies and discrepancies between different datasets and promotes fair comparison of model performance.

\begin{figure}[t]
  \centering
  \centerline{\includegraphics[width=\columnwidth]{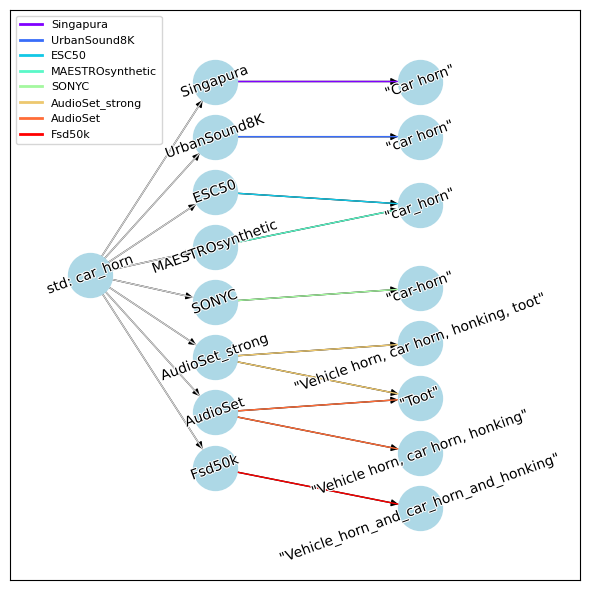}}
  \caption{Example of dataset label mapping for the standardized label \textit{car\_horn}.}
  \label{fig:dataset_label_mapping}
\end{figure}

\begin{figure}[t]
  \centering
  \centerline{\includegraphics[width=\columnwidth]{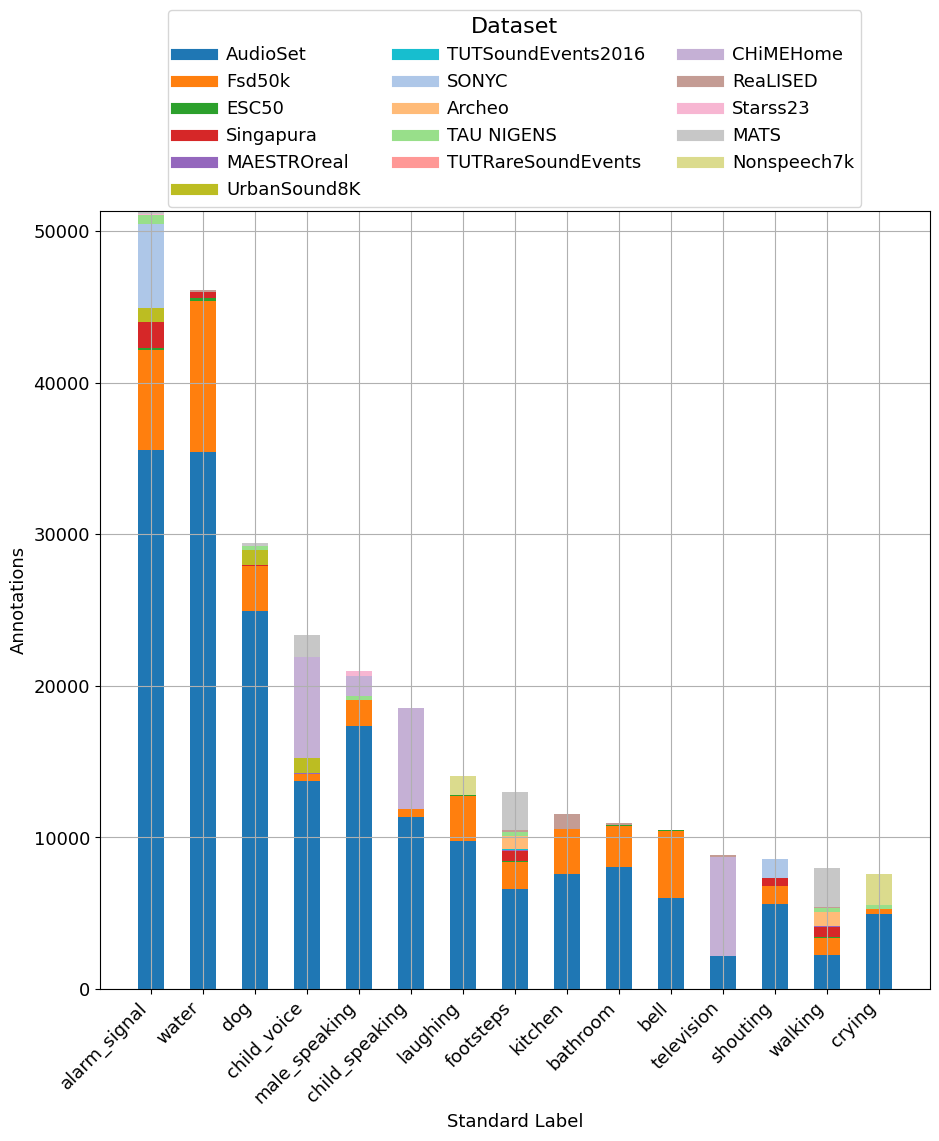}}
  \caption{The benefit of label aggregation in selected standardized labels targeting domestic sound events.}
  \label{fig:statistics}
\end{figure}

\vspace{2em}
\begin{lstlisting}[label={lst:code_example}, language=Python, aboveskip=0pt, belowskip=0pt, caption=Label search using the standardized label \textit{reverse\_beeper}, basicstyle=\ttfamily\footnotesize]
from py_salt.event_mapping import EventExplorer()

# Init taxonomy explorer
e = EventExplorer()

# Get dataset mapping dictionary
e.get_mapping_for_std_label('reverse_beeper')

{'SONYC': ['reverse-beeper'],
 'Singapura': ['Reverse beeper'],
 'AudioSet_strong': ['Reversing beeps'],
 'AudioSet': ['Reversing beeps']}
\end{lstlisting}

\section{Conclusion}
\label{section:conclusion}
In this paper, we introduced \textit{SALT: Standardized Audio event Label Taxonomy} to unify existing sound taxonomies into a global one through the standardization of labels, while also addressing some of their limitations. Built upon AudioSet's hierarchical structure, SALT standardizes and extends labels across 24 environmental sound datasets, enhancing clarity and precision and enabling cross-dataset label compatibility. Furthermore, we support the use of SALT, by introducing a Python package that provides robust tools to perform cross-dataset label aggregation, explore hierarchical relationships and visualize label mappings. These capabilities streamlining data aggregation and analysis, make SALT a valuable resource for developing machine listening systems at scale.

\section{ACKNOWLEDGMENT}
\label{sec:ack}

This work was supported by the Audible project, funded by French BPI.

\bibliographystyle{IEEEtran}
\bibliography{refs_formatted}

\end{sloppy}
\end{document}